\documentclass[conference]{IEEEtran}

\IEEEoverridecommandlockouts
\usepackage{textcomp}
\usepackage{cite}

\usepackage{graphicx}
\usepackage{lipsum}
\usepackage{amsmath} 
\usepackage{eqparbox}
\usepackage{gensymb}
\usepackage[caption=false,font=footnotesize]{subfig}
\usepackage{flushend}
\usepackage{multirow}
\usepackage{xcolor}
\usepackage{tabularx}
\usepackage{acronym}

\newacro{2x2} [2x2] {two-by-two}
\newacro{5g} [5G] {fifth-generation}
\newacro{6g} [6G] {sixth-generation}
\newacro{ADC} [ADC] {analog-to-digital converter}
\newacro{ADCs} [ADCs] {analog-to-digital converters}
\newacro{AERPAW} [AERPAW] {Aerial Experimentation and Research Platform for Advanced Wireless}
\newacro{API} [API] {application programming interface}
\newacro{awv} [AWV] {antenna weighting vector}
\newacro{BB} [BB] {baseband}
\newacro{BPSK} [BPSK] {binary phase-shift keying}
\newacro{CoMET} [CoMET] {code-modulated embedded test}
\acrodef{CP}{cyclic prefix}
\newacro{DAC} [DAC] {digital-to-analog converter}
\newacro{DACs} [DACs] {digital-to-analog converters}
\newacro{EIRP} [EIRP] {effective isotropic radiated power}
\newacro{EVK} [EVK] {evaluation kit}
\newacro{EVM} [EVM] {error vector magnitude}
\newacro{fpga} [FPGA] {field programmable gate array}
\newacro{FR2}  [FR2]  {frequency range two}
\newacro{FR3}  [FR3]  {frequency range three}
\newacro{fspl} [FSPL] {free-space path loss}
\newacro{GPIO} [GPIO] {general-purpose input/output}
\newacro{gsps} [GS/s] {giga-samples-per-second}
\newacro{I} [I] {in-phase}
\newacro{IP} [IP] {internet protocol}
\newacro{LOS} [LOS] {line-of-sight}
\newacro{LUT} [LUT] {look-up table}
\newacro{mmwave} [mmWave] {millimeter-wave}
\newacro{NSF} [NSF] {National Science Foundation}
\newacro{NR} [NR] {new radio}
\newacro{OFDM} [OFDM] {orthogonal frequency-division multiplexing}
\newacro{pcb} [PCB] {printed-circuit board}
\newacro{PPDU} [PPDU] {physical-layer protocol data unit}
\newacro{PYNQ} [PYNQ] {Python-based}
\newacro{Q} [Q] {quadrature-phase}
\newacro{QAM} [QAM] {quadrature amplitude modulation}
\newacro{RF} [RF] {radio-frequency}
\newacro{rfsoc} [RFSoC] {radio-frequency system-on-chip}
\newacro{rx} [RX] {receiver}
\newacro{SDA} [SDA]  {software-defined array}
\newacro{SDR} [SDR]  {software-defined radio}
\newacro{SNR} [SNR] {signal-to-noise ratio}
\newacro{soc} [SoC] {system-on-chip}
\newacro{TCP} [TCP] {transmission control protocol}
\newacro{TR}  [T/R] {transmit/receive}
\newacro{tx} [TX] {transmitter}
\newacro{UAS} [UAS] {unmanned aerial system}
\newacro{usb} [USB] {universal serial bus}
\newacro {usrp} [USRP] {universal software radio peripheral}
\newacro{WTR} [WTR] {waveform-triggered reception}

\usepackage{xr-hyper}
\usepackage{hyperref}
\hypersetup{
    citecolor=blue,
    colorlinks=true,
    linkcolor=blue,
    filecolor=blue,      
    urlcolor=blue,
    pdfpagemode=FullScreen,
    }
    
\begin{document}

\title{A {mmWave} Software-Defined Array Platform for Wireless Experimentation at 24-29.5~GHz} 
\author{Ashwini Pondeycherry Ganesh$^*$, Anthony Perre$^\dagger$, Alphan \c{S}ahin$^\dagger$, \.{I}smail G\"{u}ven\c{c}$^*$ and  Brian A. Floyd$^*$\\
$^*$Department of Electrical and Computer Engineering, North Carolina State University, Raleigh, NC, USA\\
$^\dagger$Department of Electrical Engineering, University of South Carolina, Columbia, SC, USA\\
e-mail:~\{apondey, iguvenc, bafloyd\}@ncsu.edu, aperre@email.sc.edu, asahin@mailbox.sc.edu
\thanks{This work has been supported by the \ac{NSF} through the award CNS-1939334 to PAWR Platform -\ac{AERPAW}.} 
}
\maketitle

\makeatletter
\def\ps@IEEEtitlepagestyle{%
  \def\@oddfoot{}%
  \def\@evenfoot{}%
}
\makeatother
\pagestyle{plain}

\begin{abstract}
Advanced millimeter-wave software-defined array (SDA) platforms, or testbeds at affordable costs and high performance are essential for the wireless community. In this paper, we present a low-cost, portable, and programmable SDA that allows for accessible research and experimentation in real time. The proposed platform is based on a 16-element phased-array transceiver operating across 24-29.5 GHz, integrated with a radio-frequency system-on-chip board that provides data conversion and baseband signal-processing capabilities. All radio-communication parameters and phased-array beam configurations are controlled through a high-level application program interface. We present measurements evaluating the beamforming and communication link performance. Our experimental results validate that the SDA has a beam scan range of -45 to +45 degrees (azimuth), a 3~dB beamwidth of 20 degrees, and support up to a throughput of 1.613~Gb/s using 64-QAM. The signal-to-noise ratio is as high as 30~dB at short-range distances when the transmit and receive beams are aligned.
\end{abstract}
% The data-rates are scalable up to 2.5~\ac{Gbps}). 
\begin{IEEEkeywords}
24-29.5~GHz, 5G, beamforming, millimeter-wave, phased array, software-defined radio, software-defined array, testbed, wireless experimentation.
\end{IEEEkeywords}

\section{Introduction}\label{Sec:Intro}
%%% Motivation and problem statement
Wireless communication at \ac{mmwave} is important for the development and implementation of \ac{5g} and \ac{6g} wireless networks that must accommodate more users with higher data rates. To evaluate the capabilities of \ac{mmwave} signals and to make optimal utilization of spectrum resources, the wireless community needs test platforms. These platforms will help power real-time applications such as autonomous vehicles and other advanced systems of the fourth industrial revolution. Therefore, there is a strong demand for affordable and high-performance \ac{mmwave} \ac{SDA} platforms for potential \ac{5g} and \ac{6g} applications above 20 GHz. 

To address these needs, multiple \ac{SDA} research efforts have recently been implemented 
%at 60~GHz but fewer have been developed 
at 28 GHz~\cite{abari_poster:_2016,Deng_VTS_2021,marinho_software-defined_2020,sadhu_128-element_2018, Sadhu_2018,wang_sdr_2019,chung_millimeter-wave_2020,XGu_2021,chen_programmable_2022,Jean_Infocom_2023}. Some implementations are stationary setups with bandwidth limitations~\cite{abari_poster:_2016,Deng_VTS_2021,marinho_software-defined_2020}; 
others are limited in beamforming capabilities and directivity~(fewer elements)~\cite{abari_poster:_2016,Deng_VTS_2021,Jean_Infocom_2023} and they do not allow for experimentation with new air interfaces using open-access softwares~\cite{abari_poster:_2016, wang_sdr_2019,Deng_VTS_2021,chung_millimeter-wave_2020}.
Some recent open-access \ac{mmwave} platforms~\cite{chen_programmable_2022,Sadhu_2018} are limited in performance due to their baseband processing rates, % and they also use expensive phased arrays. For example, 
such as \cite{chen_programmable_2022} which implements a 28 GHz mobile \ac{SDA} utilizing the \ac{usrp} 2974 that only has a bandwidth of 56 MHz. 

In this paper, we present a prototype of a cost-effective~($<$6K), open-source, reconfigurable \ac{mmwave} \ac{SDA} operating within the 24-29.5~GHz range and supporting multiple \ac{FR2} bands (e.g., n257, n258, n261). It includes a 16-element phased-array transceiver~\cite{sivers_wireless}, an AMD/Xilinx \ac{rfsoc} \ac{2x2} board~\cite{rfsoc-pynq} for baseband processing and waveform array control, and a host computer for interfacing with the test platform. Our \ac{SDA} supports 1.536~GHz bandwidth with high-speed data converters.  The open-source \ac{rfsoc} can also support an additional \ac{tx} and \ac{rx} pair for future expansion. The array provides
azimuthal beam-steering up to $\pm$45$\degree$, and 32 dBm \ac{EIRP} which can support an estimated coverage of 128~m with 20 dB link margin.
The proposed \ac{SDA} is based on affordable off-the-shelf hardware and uses a simple Python-based \ac{API}. It also enables software-based fast-beam reconfiguration of the arrays.
Altogether, this \ac{SDA} allows researchers to perform various experiments in mobile environments.
%, emulating base-station and user-equipment scenarios.
% To the best of our knowledge, this is the first low-cost, open-access, and high-performance \ac{SDA} for 24-29.5 GHz. 

%with sample rates of 4.096 and 6.554 GS/s for the \ac{ADCs} and \ac{DACs}, respectively
\begin{figure*}[t]
\centering
\includegraphics[width=0.85\textwidth, trim = 0cm 4.5cm 0cm 4cm, clip]{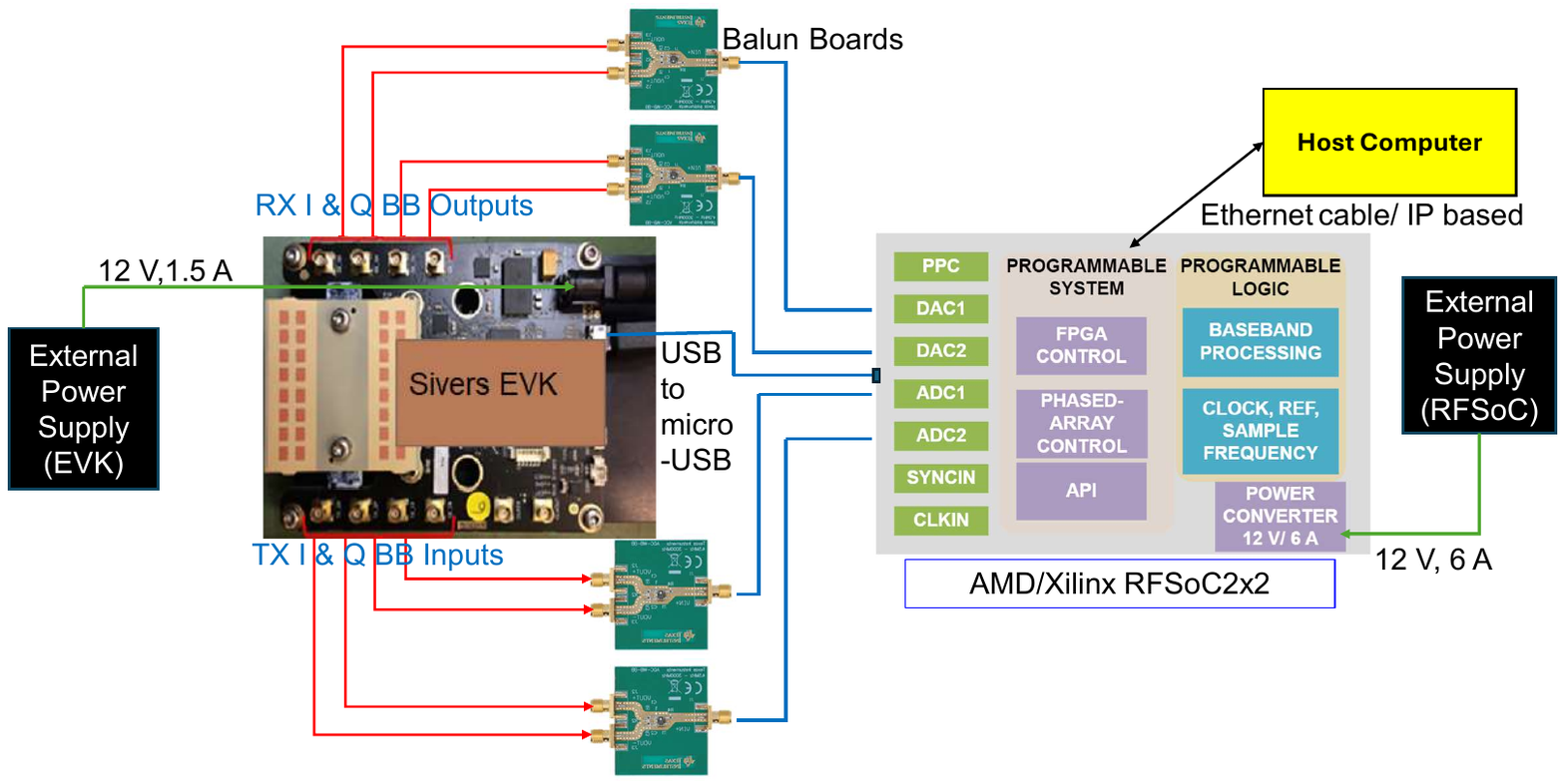}
\caption{System diagram of the SDA, including a 16-element direct-conversion phased-array transceiver evaluation kit (EVK) operating across 24-29.5 GHz and using separate antenna arrays for programmable transmit and receive beamforming;  multiple balun boards; an \ac{rfsoc}~\acs{2x2} board providing data conversion and programmable digital signal processing; a host computer; and multiple power supplies.}
\label{connections}  
\end{figure*}

The paper is organized as follows. Section~\ref{Sec:2} presents the system architecture and hardware features of the \ac{SDA}, including the phased array and its integration with the \ac{rfsoc}. Sections~\ref{Sec:3} and \ref{Sec:4} present the performance characterization measurements and sample experiment results in terms of radiation patterns, elemental gain extractions, and \ac{SNR} maps. Section~\ref{Sec:5} concludes the paper.

\section{System Architecture and Implementation}\label{Sec:2}
\subsection{Hardware Description of the SDA}
Fig.~\ref{connections} shows a block diagram of the \ac{SDA}.
It includes a) the 16-element phased-array transceiver~\cite{sivers_wireless}, b)~multiple balun boards~\cite{ti_adc-wb-bb} for differential to single-ended interfacing, c) the \ac{rfsoc} for baseband processing~\cite{rfsoc-pynq}, d) multiple power supplies, and e)~a host computer for running the application. 

The phased array is the EVK02001~\cite{sivers_wireless} from Sivers Semiconductors. The \ac{EVK} includes the transceiver chip, packaged with antennas and integrated onto a \ac{pcb} with additional circuitry for interfacing and power management. The beamforming transceiver uses direct conversion in both \ac{rx} and \ac{tx} modes to convert to/from differential analog \ac{I} and \ac{Q} \ac{BB} interfaces. 
The transceiver can be tuned from 24 to 29.5 GHz, supporting multiple \ac{FR2} bands. 

The array has 16 \ac{tx} elements and 16 \ac{rx} elements in 8x2 configurations.  Separate \ac{tx} and \ac{rx} arrays indicate that the chipset provider avoided the use of transmit/receive switches, likely to improve overall \ac{RF} performance. This comes at the penalty of increased board area for separate antennas and slight misalignment of \ac{tx} and \ac{rx} beams; however, these penalties are small at 28 GHz for an 8x2 array.
Each element within the phased array has programmable phase shifters that can be custom-defined for 64 different patterns; however, the \ac{EVK} also comes with pre-defined phase-shifter settings for 21 \textit{calibrated} patterns scanned from -45$\degree$ to +45$\degree$ with 4.5$\degree$ resolution.
The \ac{EVK} is connected to the \ac{rfsoc} via a USB interface, and control is established using an \ac{API}. 

The \ac{EVK} is responsible for the following:
    \begin{enumerate}
        \item Beamforming: Predefined or custom beam patterns are set by programming the phase-shifter values. %The principle of the weighting function is used to control the direction and the width of the beam. 
        \item Signal conditioning: The transceiver provides variable-gain amplification, filtering, and tuning.  Key metrics include a 32~dBm \ac{EIRP} for the \ac{tx} and a 6~dB noise figure for the \ac{rx}.  
        \item Frequency conversion: The \ac{RF} chain converts between \ac{BB} signals and \ac{mmwave} signals, supporting 0.9~GHz instantaneous bandwidth for both \ac{I} and \ac{Q} signals independently. 
        \item Interfacing: Analog differential  \ac{I} and \ac{Q} signals interface to the \ac{rfsoc} via baluns for differential inputs.
    \end{enumerate}

The \ac{rfsoc}~\acs{2x2} is a single-chip signal-processing platform from AMD/Xilinx~\cite{rfsoc-pynq} that integrates the key subsystems required to implement a \ac{SDR}. This includes multiple data converters to digitize \ac{BB} analog \ac{I} and \ac{Q} signals, \ac{fpga} logic operating at \ac{gsps}, and integrated processors. The \ac{rfsoc} uses a \ac{PYNQ} framework that runs on the included Arm Cortex-A53 64-bit quad processor. This \ac{PYNQ}-enabled board and its framework are used to develop our custom algorithm to integrate the RFSoC board with the phased-array platform and run operations remotely, leveraging ~\cite{mmwavesdr/design_pynq_28ghz,sahin_millimeter-wave_2023}. 

The \ac{rfsoc} is responsible for the following tasks: 
\begin{enumerate}
        \item \ac{API}: Creating an \ac{TCP}/\ac{IP} based application interface for controlling the \ac{SDA} and enabling remote-login of users for conducting experiments~\cite{mmwavesdr/design_pynq_28ghz}, thereby supporting open-source functionality. 
        \item Beam management: Programming the radio communication protocols and design of custom logic to interact with phased array via memory registers. These protocols allow one to estimate coverage and discovery delay. 
        \item Baseband signal/sample generation: Generating the continuous-time \ac{BB} \ac{I} and \ac{Q} signals to be transmitted by the \ac{EVK} using the baseband samples provided by the host computer or providing the baseband samples by sampling the continuous-time \ac{I} and \ac{Q} signals. 
\end{enumerate}

 % Data processing, packet generation, and modulation/~demodulation are handled by the host computer. More details on the waveform and achievable data rates are provided in Section~\ref{PPDU}.

\subsection{RF Link Budget}
We calculate the range ($R$) as 128~m for a given \ac{EIRP} of 32~dBm at bore-sight, using the following formula~\cite{Skolnik_Radar_2008}:
\begin{equation}
\begin{aligned}
R &= \frac{\lambda}{4\pi}\sqrt{\frac{P_{\rm EIRP} \cdot G_r}{P_{\text{sens}} + P_{\rm LM} + L_a}}~,
\end{aligned}
\end{equation} where $G_r$ is receive antenna gain, \( \lambda \) is the wavelength of the signal, \( P_{\text{sens}} \) is receiver sensitivity, $P_{\rm LM}$ is link margin~(here set to 20~dB to accommodate additional losses due to implementation errors), and $L_a$ is atmospheric losses. The \ac{fspl} for this range is 104~dB, determined by:
%\vspace{-0.2cm}
\begin{equation}
\begin{aligned}
        \text{FSPL} = 20 \log_{10}\left(\frac{4 \pi R}{\lambda}\right).
\end{aligned}
\end{equation} 
Reduced wavelength leads to higher path loss, hence higher gain antennas are desirable to increase the range.
%\vspace{-0.04cm}
\subsection{Beam-Pattern Control}  
The control of the phased array is done dynamically by programming the response of each array element to realize a specific phase shift to form beam patterns according to the standard phased-array theory below: 
\begin{equation}\label{AF}
AF(\theta) = \sum_{n=0}^{N-1} A_n \cdot e^{j 2\pi n \frac{d}{\lambda} \sin \theta -\phi_n} ,
\end{equation} 
where $\theta$ is the angle of departure/arrival with respect to the broadside, $d$ is the distance between the elements, and where $A_n$ and $\phi_n$ are the amplitude and phase of the  \textit{n}\textsuperscript{th} element.
For each desired pattern, a user-defined \ac{awv} is generated for all phase shifters within the array. 
These are vector interpolators that realize an element response of
$ Z_n = A_n \cdot e^{j\phi_n} =I_n+jQ_n$, %\label{eq:Z}
%in a Cartesian format, 
as follows:
\begin{align}
I_n &= A_n \cdot \cos(\phi_n), & Q_n &= A_n \cdot \sin(\phi_n)~, \label{eq:IQ}
\end{align}
where \ac{I} and \ac{Q} components are the in-phase and quadrature-phase response of the \textit{n}\textsuperscript{th} element, controlled with amplifiers. 

\subsection{Software Control and Functionality}
The software control is partitioned into three categories, each allowing control of a set of functions. The top layer is the scripts running on the host computer. The \ac{API} is created on the \ac{rfsoc}'s \ac{PYNQ} framework, which serves as an interface to control the hardware via a \ac{TCP}/\Ac{IP}-based network. The \ac{API} enables a high-level execution of control commands that are called from top-level scripts. The second and major part of the functionality is handled by the \ac{rfsoc} board. We set custom \ac{IP} addresses for the \ac{fpga} boards to access the boards and communicate. The \ac{rfsoc} controls the phased array via a USB interface. The third part is \ac{EVK} control, that includes \ac{tx}/\ac{rx} mode of operation, variable gain setting, phase of antenna elements, number of active elements at a given time, and beamforming control. The partitioning of functions between the servers and the \ac{fpga} enables fast execution and flexibility of remote experiments. More details on software mapping of the prototype can be found in~\cite{sahin_millimeter-wave_2023}.

\subsection{PPDU Design}\label{PPDU}
\begin{figure}[t]
\centering
\includegraphics[width=3.2in, trim = 0.0cm 0cm 0cm 0cm, clip]{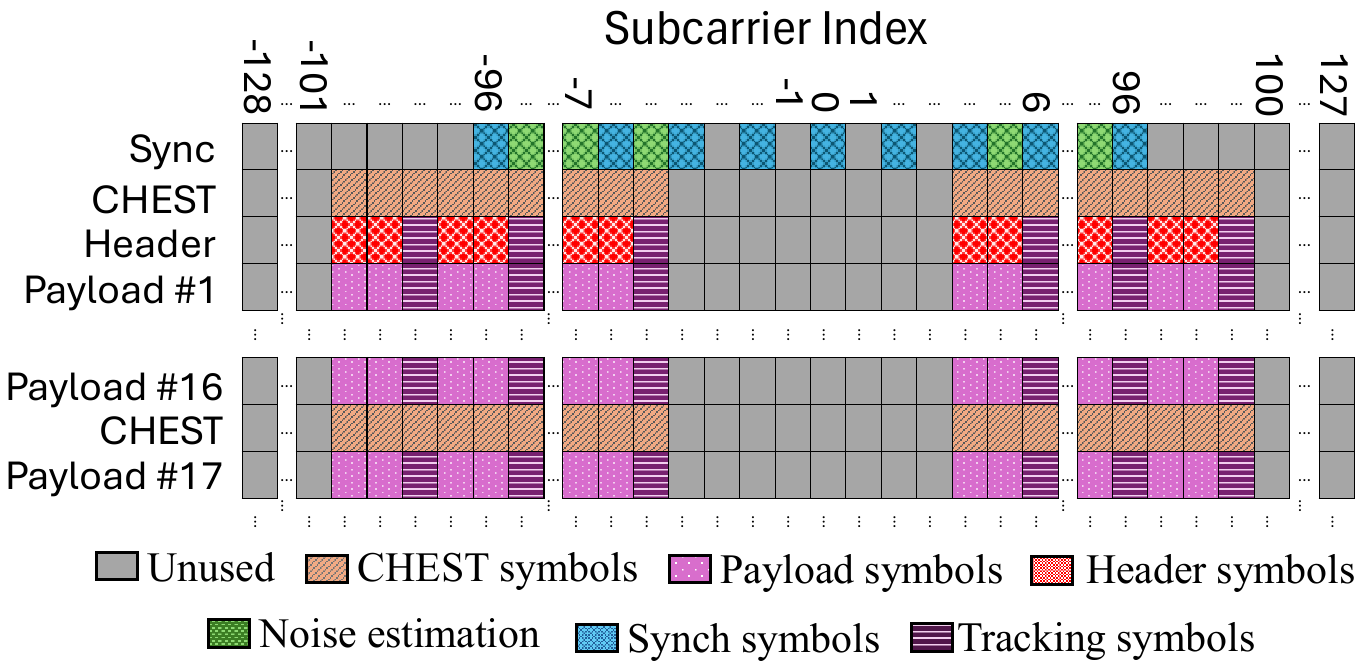}
\caption{Physical-layer protocol data unit structure and subcarrier mapping.}
\label{Subcarrier Layout}  
\end{figure}

We adopt the \ac{OFDM}-based \ac{PPDU} design discussed in~\cite{sahin_millimeter-wave_2023} and increase the throughput by supporting higher-order modulation. Baseband processing, channel estimation, and protocol implementation are done using key technologies such as \ac{WTR}. As illustrated in Fig.~\ref{Subcarrier Layout}, the transmitter encodes data bits along with a preamble for time-frequency synchronization and channel estimation at the receiver. The channel response is estimated based on preamble symbols. The time-domain signals are converted to a baseband frequency domain signal using the \ac{DACs} on the \ac{rfsoc}. On the receiver end, the decoding process involves synchronization, estimating the non-idealities such as carrier-frequency offset in the received signal, removing the \ac{CP}, estimating the channel parameters, and extracting the payload data. 
% A synchronization waveform based on a Golay sequence is used as a reference signal for time and frequency synchronization. 

The \ac{PPDU} supports BPSK, 4-QAM, 16-QAM, and 64-QAM with Gray mapping. Depending on the modulation type, the number of payload bits may not be an integer multiple of the symbol bit length; additionally, the remaining payload bits for the last codeword may not be enough to completely fill it. To resolve these issues, the payload bits are pre-padded with zeroes to fill the remaining modulation symbols and codewords. The payload bits are then post-padded with zeros so that each data subcarrier in the last OFDM payload symbol is filled. All padding information is conveyed in the header field of the \ac{PPDU}. The payload OFDM symbols use $N_{\text{sub}}=128$ active data subcarriers, where the IDFT size is 256 and the \ac{CP} size is 64, with an encoding scheme that features $L_{\text{codeword}}=128$-bit codeword. Additionally, the channel encoder uses $r_{\text{code}}=1/2 $ polar code and $L_{\text{crc}}=8$  cyclic-redundancy check bits are appended to message bits per codeword. An extra chest field occurs every 16 OFDM payload symbols and is used to refresh the channel estimate. Tracking symbols are also used to estimate the common phase error. 

The bandwidth of the complex baseband signal can be calculated as $f_\text{s}\times$(192 active + 8 DC tones)/256, i.e., $1.2$~GHz, where $f_\text{s}= 1.536$~Gsps is the sample rate of the \ac{SDA}.  The data rate of the payload can also be calculated as
\begin{equation}
\begin{aligned}
{R_{\text{data}}} &= \frac{N_{\text{sub}} \cdot \log_{2}\left({M}\right)}{T_{\text{sym}}}\left({r_{\text{code}} - \frac{L_{\text{crc}}}{L_{\text{codeword}}}}\right)~,
\end{aligned}
\end{equation} 
where $M$ is modulation order and $ T_{\text{sym}}=208.3$~ns  is OFDM symbol duration, including the \ac{CP}.  For instance, the data rates for BPSK and 64-QAM are given by  $268.8$~Mbps and $1.613$~Gbps, respectively.
% be obtained as $56$~bits and $448$~bits over $T_{\text{sym}}=208.3$~ns, i.e.,

\section{Beam Steering Characterization}\label{Sec:3} 
\begin{figure}[t]
\centering
\includegraphics[width=\columnwidth, trim = 0cm 3cm 0cm 3cm, clip]{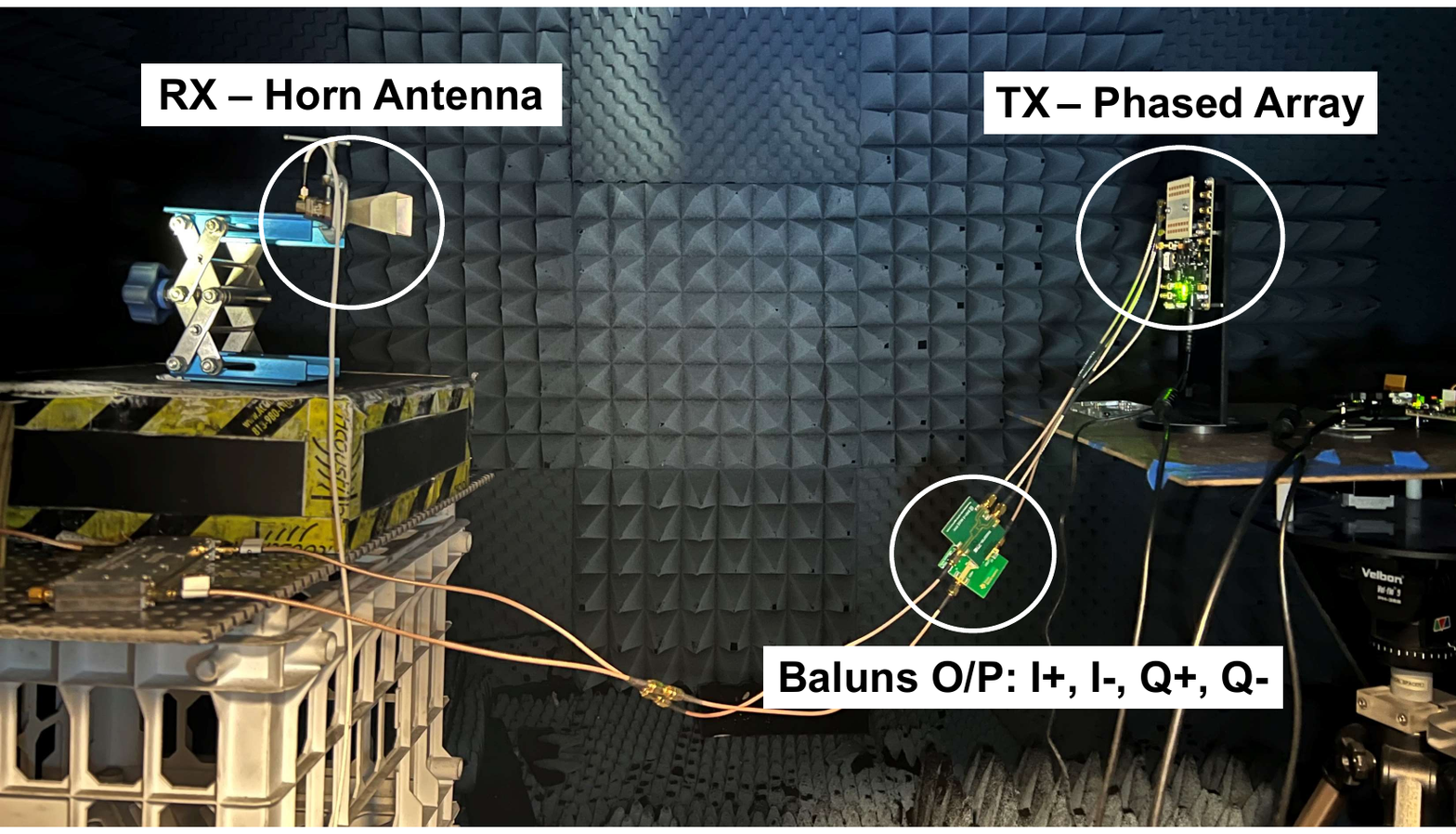}
\caption{Over-the-air transmit mode measurement setup, using the phased-array evaluation kit in \ac{tx} mode and a horn antenna connected to a spectrum analyzer as the \ac{rx}.}
\label{TX-mode_test__setup}  
\end{figure}

This first set of experiments focuses on understanding the beamforming response of the phased array by evaluating its antenna patterns. %azimuthal direction from -45$\degree$ to +45$\degree$. 
Fig.~\ref{TX-mode_test__setup} shows the measurement setup inside an anechoic chamber for evaluating the over-the-air performance of the array in transmit mode. We use the \ac{EVK} as the \ac{tx} and inject an 800 MHz continuous wave (complex) signal into the baseband of the \ac{tx} to generate a single-sideband tone at 28.8~GHz. A standard gain horn antenna connected to a spectrum analyzer is placed at a 1.3~m distance and used as the \ac{rx}. The broadside pattern is first measured while rotating the \ac{EVK} along the azimuth.
%are derived from measurements outlined in Section \ref{Array_response}. 

\begin{figure}[t]
\centering
\includegraphics[width=\columnwidth]
{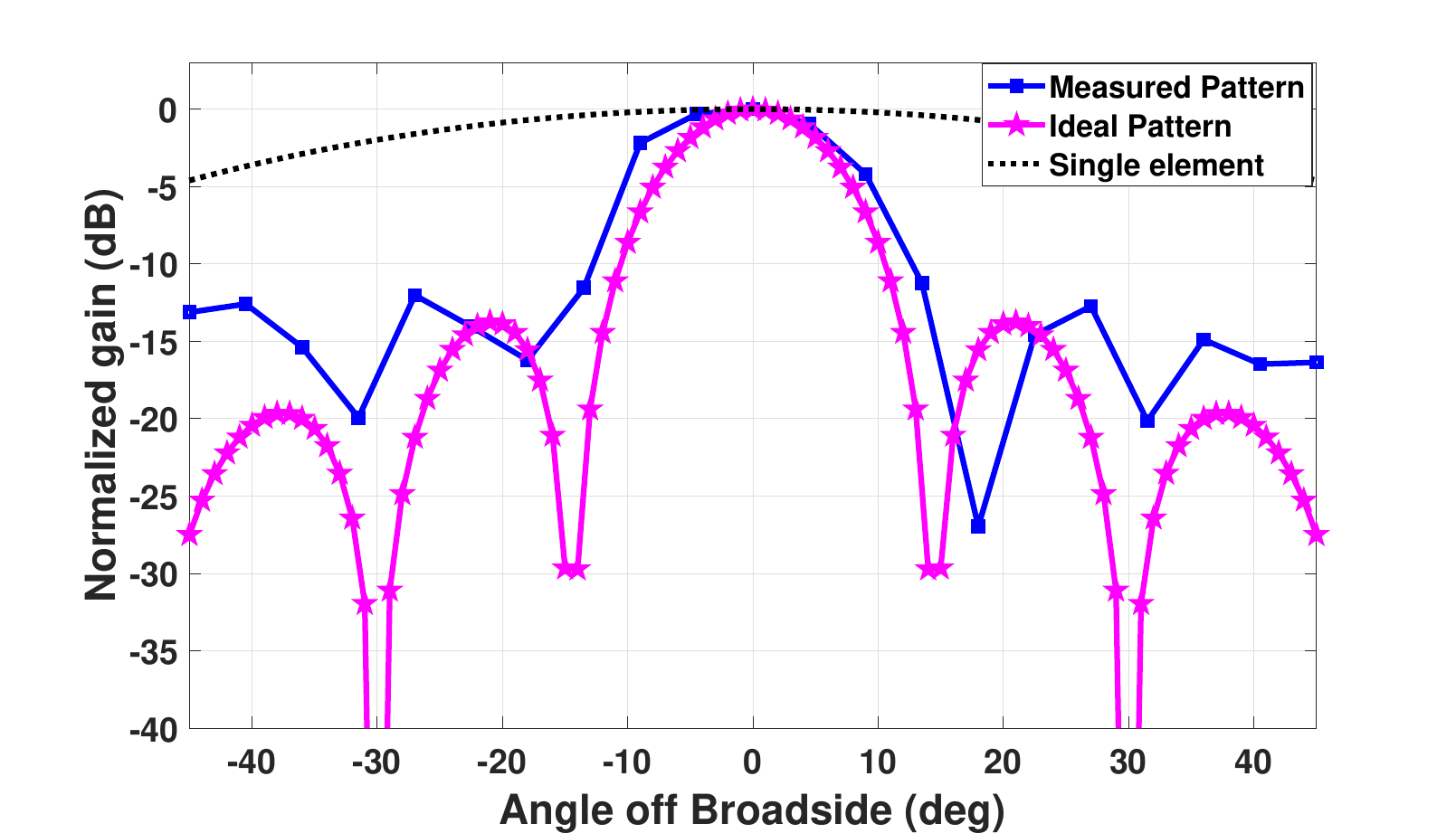}
\caption{Measured versus ideal beam pattern for an 8x2 \ac{tx} array steered to broadside together with single-element response.}
\label{ES Pattern}  
\end{figure}

\begin{figure}[h]
\centering
{\includegraphics[width=\columnwidth,height=.3\textwidth,  trim = 0cm 0cm 0cm 1cm, clip]{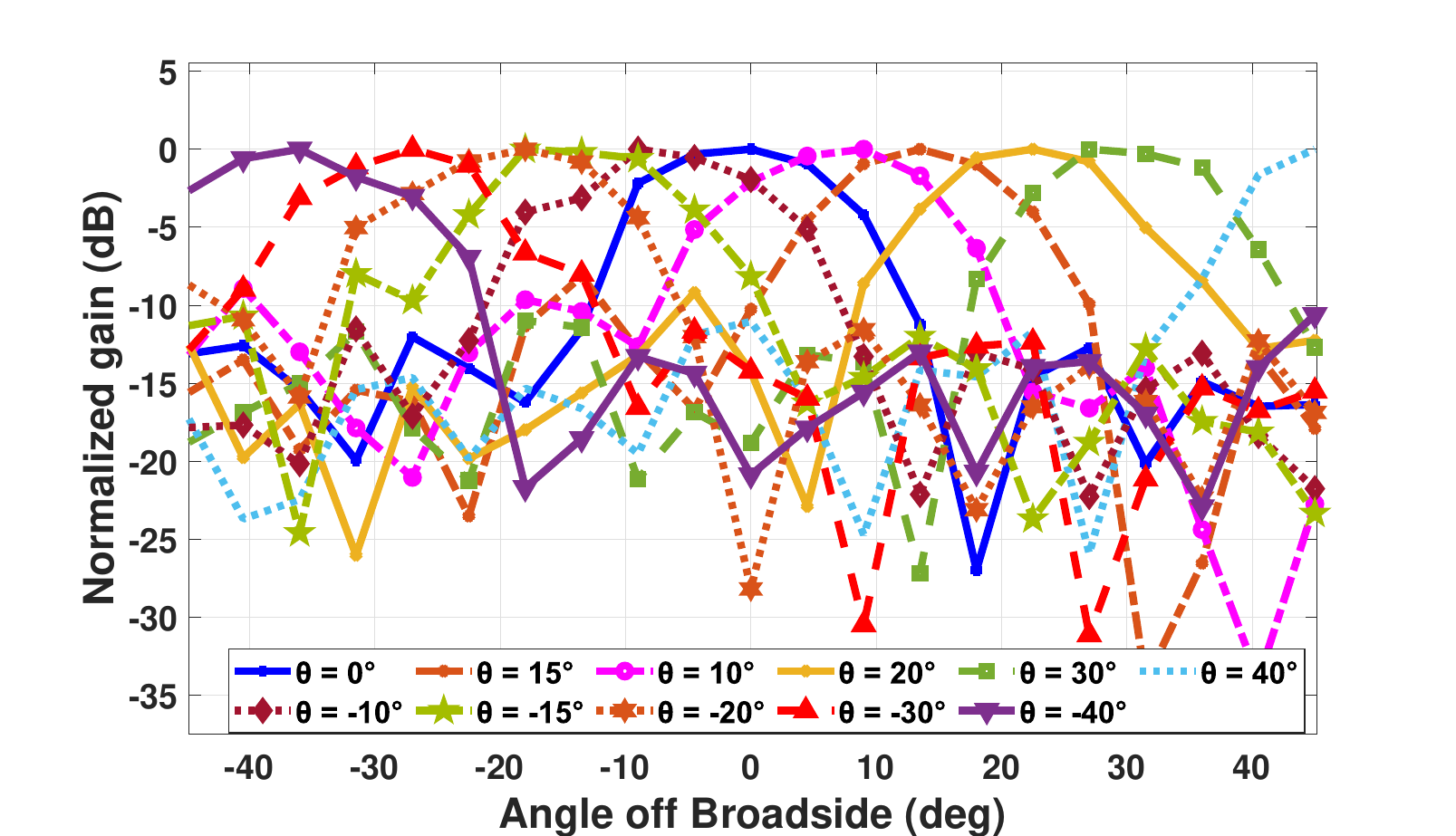}}
\caption{Measured beam patterns for TX array, when steered across $\pm$45$\degree$.}
\label{MS Pattern}  
\end{figure}

\begin{figure}[t!]
\centering
\includegraphics[width=\columnwidth,height=.29\textwidth]{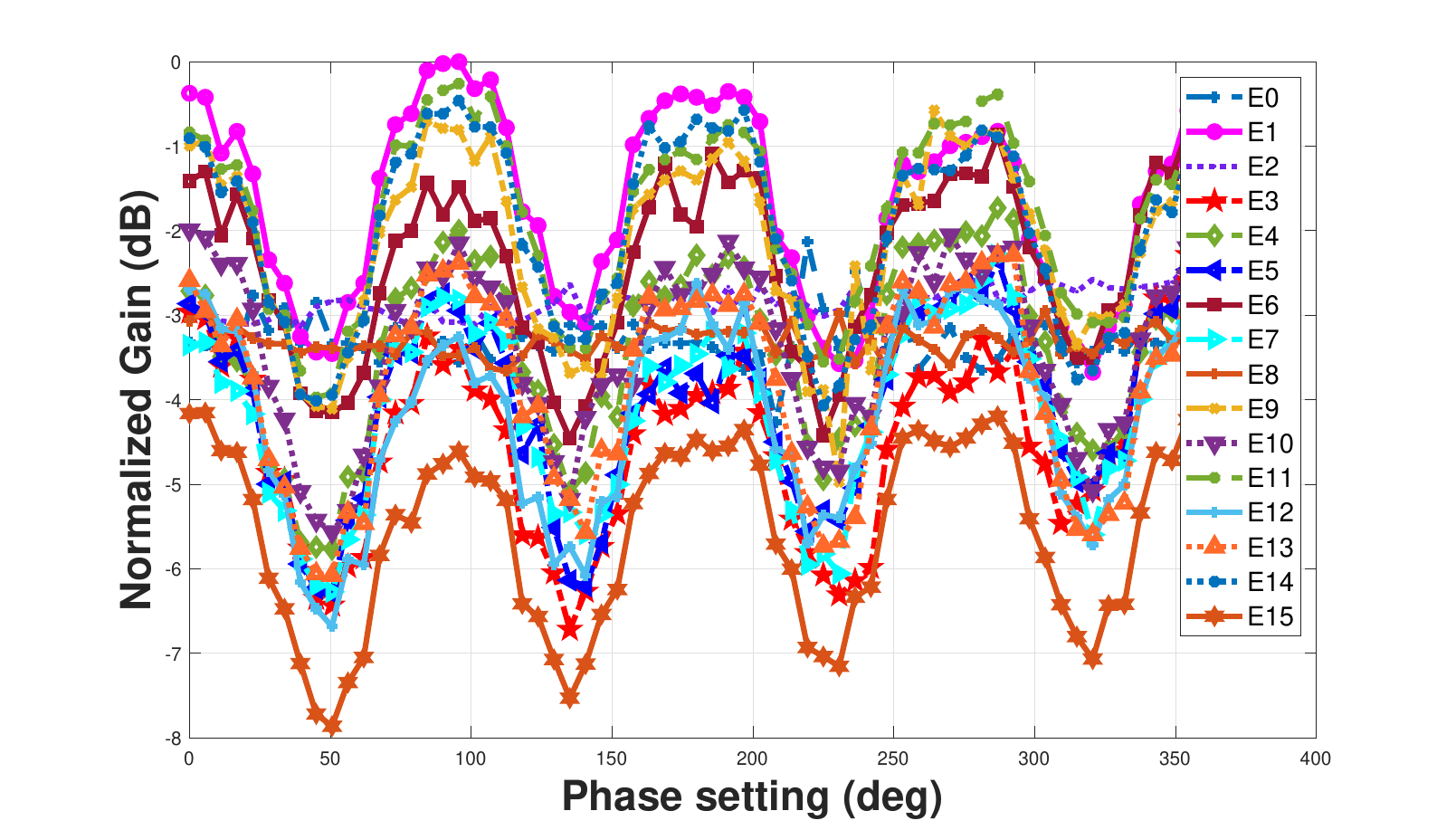}
\caption{Gain for each element of the array in transmit mode, extracted using the CoMET built-in self-test technique \cite{9975103}.}
\label{Gain}  
\end{figure}

Fig.~\ref{ES Pattern} shows the array pattern normalized to a peak gain of 14 dBi. %when generating the broadside pattern using the manufacturer-provided \ac{awv}. 
It is compared to the ideal response of an 8x2 array and an ideal unit-element antenna response with a $\cos(\theta)$ pattern. As can be seen, there is a notable alignment in the patterns. However, the side lobes are slightly higher than desired. The measured peak-to-null ratio is 17~dB, which indicates that the \ac{awv} is not perfectly calibrated. The measured 3~dB beamwidth is $\pm$10$\degree$.

Fig.~\ref{MS Pattern} shows the measured antenna patterns as the beams are steered from -45$\degree$ and +45$\degree$.  We do not observe any unexpected behavior as the beam is steered; however, we also conclude that better performance, in terms of null depth and side-lobe levels, could be obtained with a better calibrated \ac{awv} for these patterns, also stated in~\cite{Deng_VTS_2021}. Note that the manufacturer only provides pre-defined \ac{awv}s for the $\pm$45$\degree$ range, presumably due to mutual coupling between antennas limiting the performance outside of this range.

Our ultimate goal is to enable arbitrary pattern formation, which necessitates the evaluation and calibration of each individual element in the \ac{EVK}. Calibration is crucial to equalize gain and phase responses, ensuring uniform performance regardless of antenna configurations. To this end, we extracted the element-by-element response using a built-in self-test technique for phased arrays developed at NC State. This \ac{CoMET} technique uses orthogonal on/off codes \cite{9975103} per element in the array to modulate a test signal, and these signals are then code multiplexed in free space for \ac{tx} mode. A simple power detector measures the aggregate response and, in so doing, squares the signal and creates all possible cross-products (or correlations) between elements, each modulated by code products. The codes are selected to have \textit{orthogonal} code products; hence, all correlations can be extracted. These can be used within a transcendental equation solver in MATLAB to determine each element's gain and phase response. More details can be found in \cite{Hong_TMTT_2020,9975103}. 

The \acp{awv} provided by the manufacturer does not provide calibrated phase settings for all possible phase shifts, but rather settings for particular calibrated \textit{patterns}. As such, we programmed each phase shifter assuming they had an ideal Cartesian response from (\ref{eq:IQ}), using the \ac{DACs} for the \ac{I} and \ac{Q} settings in each phase shifter.

Fig.~\ref{Gain} shows the extracted gain versus ideal phase setting. %, where the \ac{DACs} controlling the vector interpolating phase shifter within each element are controlled according to Equation~\eqref{eq:IQ}. 
We observe that the gain fluctuates in a periodic fashion, having maximum gain near the 0/90/180/270$\degree$ axes, and minimum gain in between. This is not unexpected, as vector interpolators can have such variations, which is why they must typically be calibrated.  The other key observation is that the element-to-element variation is within 4-5 dB. These variations can be due to chip, package, and/or antenna variations between elements. Future work for our team is to fully calibrate the \ac{EVK} using \ac{CoMET} for all possible phase settings, which will allow the synthesis of arbitrary patterns.

\section{Communication Link Measurements}\label{Sec:4}
\begin{figure}[t]
\centering
  % {\includegraphics[width=.23\textwidth,height=.15\textwidth]{reflector product outline.png}}\quad
\includegraphics[width=\columnwidth, trim = 1cm 9cm 0cm 7.5cm, clip]{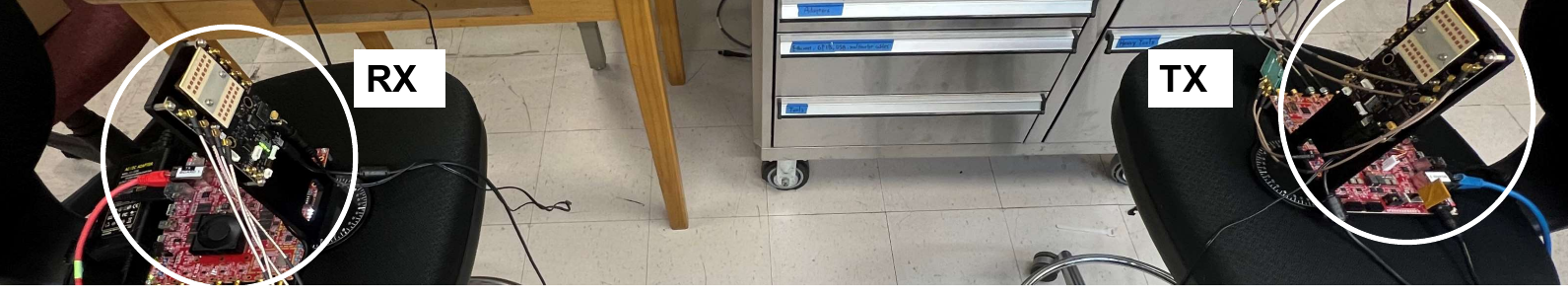}
\caption{Beam-sweeping experiment setup with transmitter and receiver software-defined arrays (SDAs) placed at a distance of 4.5~m in an indoor environment. The distance in the image is not to scale.}
\label{beam_sweep}  
\end{figure}

In this experiment, we use the \ac{SDA} and validate the link performance for multiple beam patterns.
%when transferring 0.3 Gbps data. %by conducting experiments to test and characterize the RF link established at close range for a variety of beam alignments when transmitting 1 Gbps data.
The measurements are performed in an indoor environment using a tabletop setup, as shown in Fig.~\ref{beam_sweep}. The two \ac{SDR}s~(\ac{tx} and \ac{rx}) are placed at a distance of 4.5~m and are controlled by a companion computer over an access point through a \ac{TCP}/\ac{IP} based connection. The beam index information is transmitted, as the \ac{rx} has no information about the beam index at the \ac{tx}. This experiment uses predefined beams and electronically sweeps these in the azimuthal direction from -45\degree~to +45\degree.

The transmitter and the receiver have their own set of 21~\acp{awv}. The \ac{tx} will transmit the beam at each position for a specific time and then change the direction in the next time frame until all scan angles from -45$\degree$ to +45$\degree$.
\begin{figure}[t]
\centering
\subfloat[\ac{tx} steered to -45$\degree$.]{\includegraphics[width=1.6in]{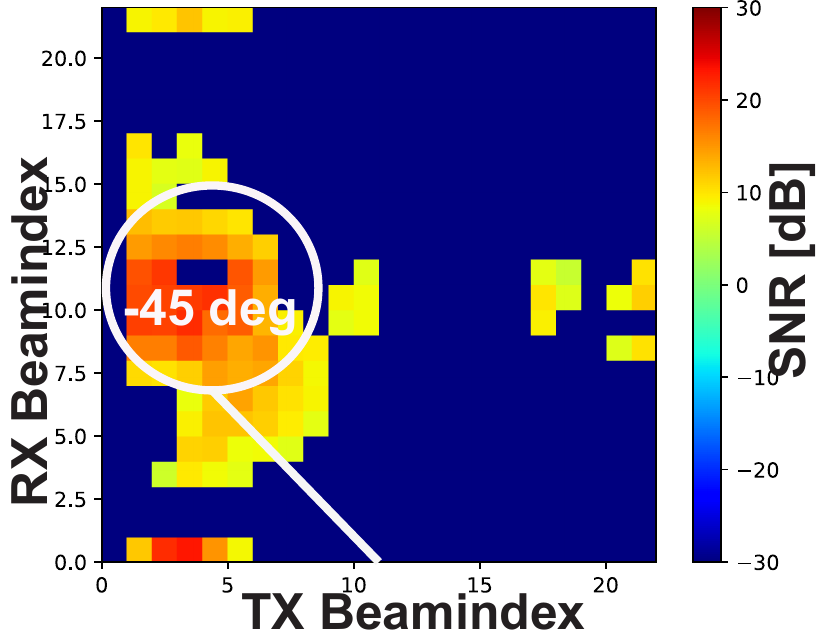}\label{fig:sub1}}
\subfloat[TX steered to +45$\degree$.]{\includegraphics[width=1.6in]{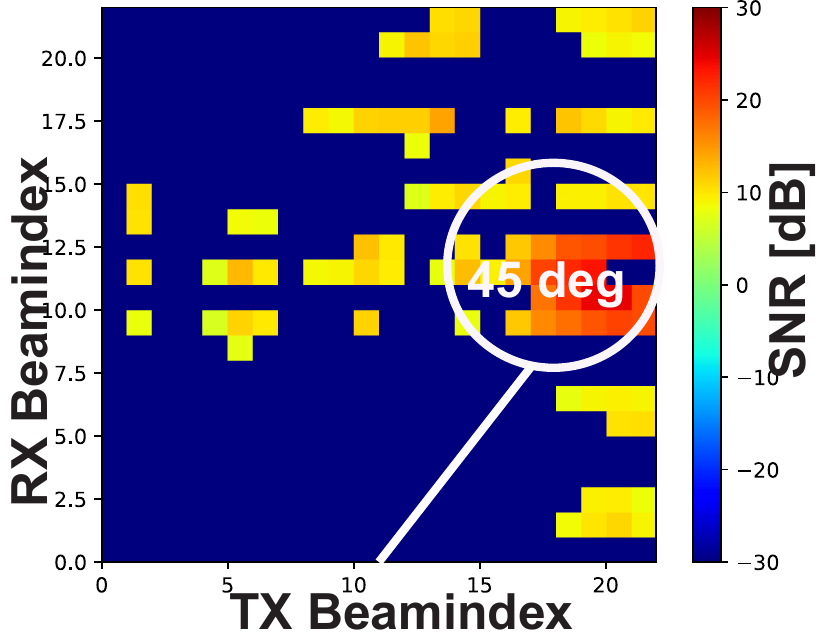}\label{fig:sub2}}\\
\subfloat[\ac{tx}-\ac{rx} in line-of-sight~(LOS).]{\includegraphics[trim=0 45 0 0, width=1.6in]{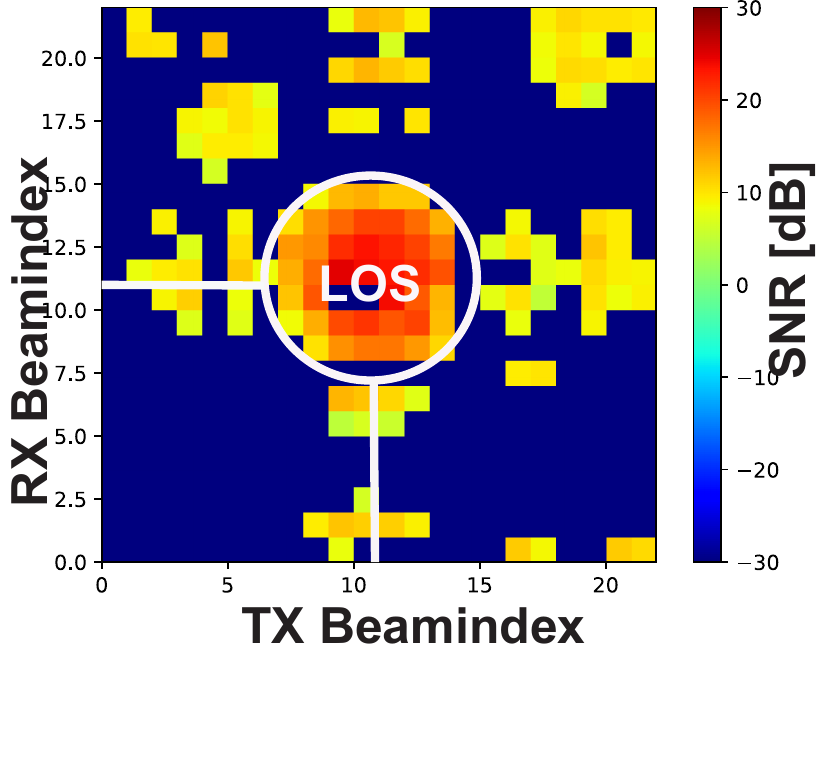}\label{fig:sub3}}	 \caption{Measured \ac{SNR} of the received signal as both \ac{tx} and \ac{rx} beams are swept across full range.}
\label{fig:SNR Pattern}
\end{figure}
Using this measurement, we evaluated the \ac{SNR} for each beam pattern on the transmit and receive sides, the data transfer rates, and the quality of the data link in terms of \ac{EVM}. The \ac{SNR} of the received signal is shown in Fig.~\ref{fig:SNR Pattern} across all swept patterns. Each beam-index position corresponds to a scan angle, with 0$\degree$ at the center for beam-index = 11. The matrix demonstrates that the \ac{SNR} reaches its peak of 30~dB when the \ac{tx} and \ac{rx} beams are perfectly aligned, specifically at \ac{LOS}. %\ac{tx}-\ac{rx} beam-index~(11). 
From this \ac{SNR}, we note that the \ac{EVM} of the system must also be -30 dB or lower, consistent with data in \cite{sivers_wireless}. We also see a minimum \ac{SNR} of 10~dB across all linear sweep positions. 

Note that the measurement setup in Fig.~\ref{beam_sweep}, is a real-time scenario susceptible to multiple sources of reflections to maintain a link. Notably, a reflected signal is observed at the convergence of \ac{rx} beam-indices~(15-18) and \ac{tx} beam-indices~(3-8) in Fig.~\ref{fig:SNR Pattern}\subref{fig:sub3}, attributed to the presence of a metal cabinet within the setup. Fig.~\ref{fig:SNR Pattern}\subref{fig:sub1}-\ref{fig:SNR Pattern}\subref{fig:sub2}, refer to the SNR pattern when \ac{SDA} is steered to -45\degree~and +45\degree, respectively.

\begin{figure}[t]
\centering
\subfloat[4~QAM]{\includegraphics[width=1.16in, trim = {1.5cm 0cm 1.5cm 0cm},clip]{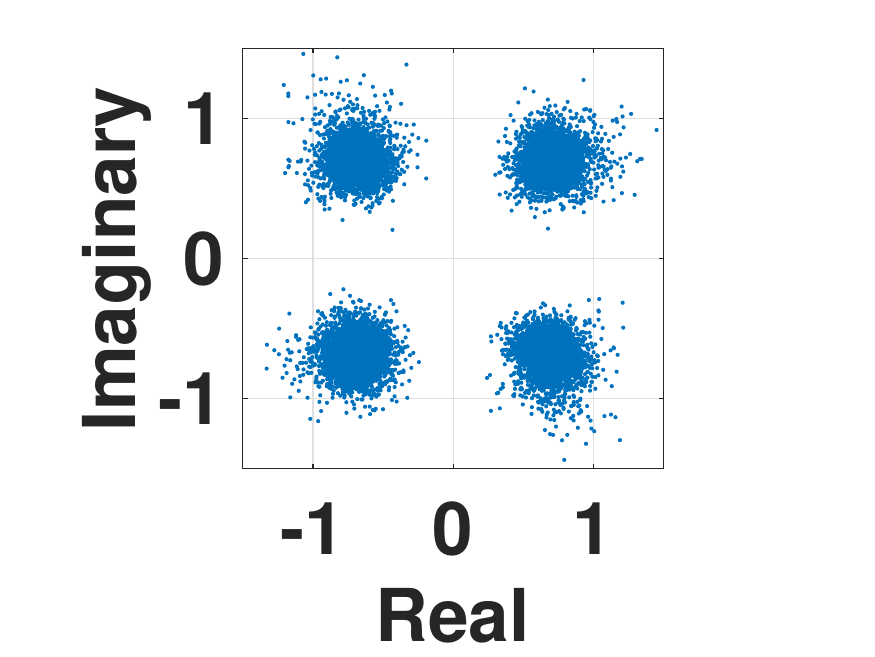}\label{subfig:qpsk}}
\subfloat[16~QAM]{\includegraphics[width=1.16in, trim = {1.5cm 0cm 1.5cm 0cm},clip]{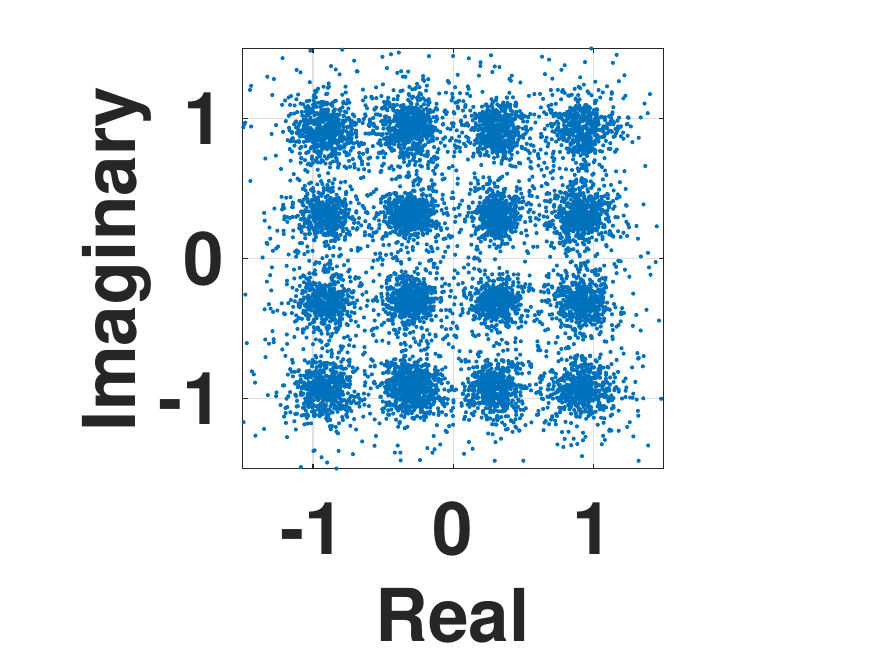}\label{subfig:16qam}}
\subfloat[64~QAM]{\includegraphics[width=1.16in, trim = {1.5cm 0cm 1.5cm 0cm},clip]{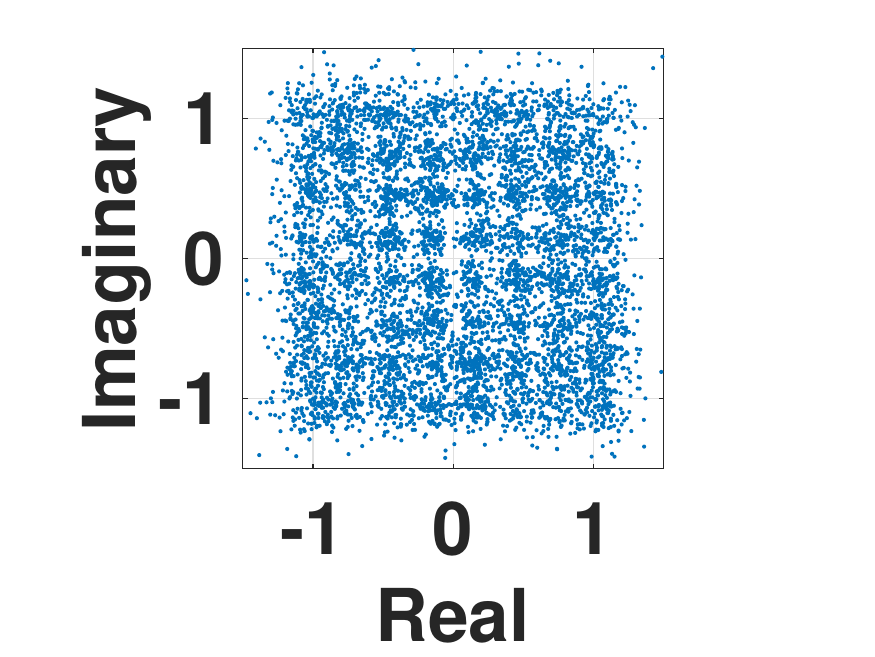}\label{subfig:64qam}}  
\caption{The received symbols for 4~QAM, 16~QAM, and 64~QAM.}
\label{fig:constellationPlots}
\end{figure}
Fig.~\ref{fig:constellationPlots} shows the received symbols plots for 4-QAM, 16-QAM, and 64-QAM after channel equalization. Here, we test our \ac{SDA} by transmitting an ASCII message as a payload with QAM constellations. In each case, we calculate the log-likelihoods of the coded bits for different constellations, and its corresponding codewords are decoded. For each modulation, the transmitted ASCII message is recovered successfully. In our setup, we observe that some of the subcarriers can fade approximately by 3-6 dB (slight frequency-selective fading within 1.2 GHz bandwidth), which causes noisy symbols in Fig.~\ref{fig:constellationPlots}. In the future, we plan to investigate the channel characteristics by using our SDA to improve the reliability of the link. Further improvements to our design may also include higher-order modulations along with different coding strategies, which would further increase the maximum data rate of the system. We also plan to configure these boards with WiFi instead of ethernet cables, allowing us to perform outdoor tests~($>$10~m) to unleash its full potential.

\begin{table*}
\caption{Performance Summary and Comparison with Existing Solutions.}
\centering
\small
\renewcommand{\arraystretch}{1.1}
\begin{tabularx}{\textwidth}{|c|c|c|c|c|c|c|c|c|}\hline
       {Reference}& 
       {\cite{abari_poster:_2016}}& 
       {\cite{sadhu_128-element_2018,Sadhu_2018,XGu_2021,chen_programmable_2022}}&
       {\cite{wang_sdr_2019}}& 
       {\cite{marinho_software-defined_2020}}&
       {\cite{chung_millimeter-wave_2020}}&
      {\cite{Deng_VTS_2021}}&
       {\cite{Jean_Infocom_2023}}&
       {\textbf{This work}}\\ 
       \hline  
       % \textbf{Venue} & ACM 2016~(MIT)&&&&&&& \\
       % \hline
       {Frequency}& 24~GHz& 28~GHz & 28~GHz& 28~GHz& 28~GHz&28~GHz & 26.5-40~GHz& \textbf{24-29.5~GHz} \\ 
       \hline
       {Bandwidth} & 200~MHz & 100~MHz & 100~MHz &NR& 20~MHz & 40~MHz & 20~MHz  & \textbf{1.2~GHz}\\
       \hline
       % \multirow{3}{*}{\textbf{Number of elements}} & $<$10 for MIMO\\ &Upto 128 elements for mmWave \ &16 elements\\
       {Elements} & 8 & 128&NR&8 & 64& 8 &horn&16\\
       \hline
       {DSP} & X310 &2974/B200 & RIO& N310& NI 294xR & M3-Force& N210  & \textbf{\ac{rfsoc}~\acs{2x2}} \\
       \hline
       {Range}&$>$100 m& $>$100 & 7.1 m&NR&NR&NR& 0.7 m& \textbf{128 m}\\
       \hline
       {Beamwidth} &NR&NR&NR&27\degree &NR&12\degree& 26\degree  & 20\degree\\
       \hline
       {Polarization}& Single & Dual & Single & Single& Dual &NR& Single  & Single \\
       \hline
       {Beamforming}&Analog &Hybrid& Analog &Digital& Hybrid & NA &  NA&Analog\\
       \hline
        {Power}&NR&NR&NR&NR&NR&NR&NR& \textbf{5.5 W}\\
       \hline
        {Control} &{GNU radio} &{API} & LabVIEW & MATLAB &LabVIEW & MATLAB & GNU radio & \textbf{\ac{API}}\\
        % \multirow{3}{*}{}
        % &  \textcolor{blue}{Open-access} \\
       \hline
       {Interface}&GigE & GigE&GigE &GigE&GigE &GigE &  GigE&  \textbf{GigE/USB/SPI} \\
       \hline
        {Data rate}&NR&NR&400~Mbps&NR&NR&NR&NR& \textbf{1.613 Gbps} \\
       \hline
\end{tabularx}
\label{tab:performance comparison}
\vspace{1mm}
    \begin{minipage}{\textwidth}
        \centering
        \footnotesize
        NA - Not Applicable, NR - Not Reported.
    \end{minipage}
\end{table*}

% ==================
% # Conclusion #
% ==================
\section{Conclusion}\label{Sec:5}
In this work, we present a low-cost \ac{SDA} platform. To the best of our knowledge, this is the highest capability and lowest cost open-access \ac{SDA} for 24-29.5 GHz. Table~\ref{tab:performance comparison} summarizes details of performance achieved, compared to existing solutions. This prototype can be used to test new wireless protocols for \ac{5g} and advanced \ac{6g} systems that will support artificial intelligence and automation applications. We test our developed system by establishing an \ac{RF} link and characterizing its responses. We evaluated the signal strength received in the presence of multipath components and the array's response in an over-the-air free-space environment. The received signal strength is highest with an SNR of 30~dB when the transmit and receive beams are aligned with each other. The array beam response shows near-expected side-lobe levels, 3~dB beamwidth of $\pm$10$\degree$ and a scan range of $\pm$45$\degree$. 

The developed \ac{SDA} can be used by researchers to conduct experiments using programmable \ac{mmwave} radios in real-world environments. This includes waveform development, beamforming algorithms, spectrum monitoring, surveillance, and interference detection. Future work includes completing the calibration of the array for arbitrary pattern synthesis and mounting the \ac{SDA} to an \ac{UAS} for mobile experiments within the \ac{AERPAW} platform.
%, increasing the data rates using higher density modulation schemes, 

%Extending the scan angle from -90$\degree$ to +90$\degree$ is challenging due to mutual coupling between the elements, lack of data on inherent systematic polarization between the elements, and the transmission trace lengths used. 
 %applications in test fields to understand spatial diversity and support future research on \ac{6g} and beyond with the existing scan range. It can enable signal and

\section{Acknowledgements}
The authors thank Prof. Suresh Venkatesh from NC State University for technical discussions and feedback. 

% ==============
% # REFERENCES #
% ==============
% \bibliographystyle{IEEEtran}
% \bibliography{IEEEabrv,biblio_rectifier}
%\bibliographystyle{IEEEtran}
% \bibliographystyle{myIEEEtran.bst}
\bibliographystyle{IEEEtran}
\bibliography{IEEEabrv,references,floyd_refs}
\end{document}